\documentclass[twocolumn,twocolumn]{IEEEtran}
\usepackage[T1]{fontenc}
\usepackage{color}
\usepackage{float}
\usepackage{mathrsfs}
\usepackage{amsmath}
\usepackage{amssymb}
\usepackage{graphicx}
\usepackage[unicode=true,
bookmarks=true,bookmarksnumbered=true,bookmarksopen=true,bookmarksopenlevel=1,
breaklinks=false,pdfborder={0 0 0},pdfborderstyle={},backref=false,colorlinks=false]
{hyperref}
\hypersetup{pdftitle={Your Title},
	pdfauthor={Your Name},
	pdfpagelayout=OneColumn, pdfnewwindow=true, pdfstartview=XYZ, plainpages=false}

\makeatletter

\floatstyle{ruled}
\newfloat{algorithm}{tbp}{loa}
\providecommand{\algorithmname}{Algorithm}
\floatname{algorithm}{\protect\algorithmname}

\usepackage[caption=false,font=footnotesize]{subfig}
\usepackage{algorithm}
\usepackage{algorithmic}

\usepackage{multirow} 
\usepackage{amsmath} 
\usepackage{xcolor}

\allowdisplaybreaks[4]

\ifCLASSOPTIONcompsoc
\usepackage[caption=false,font=normalsize,labelfont=sf,textfont=sf]{subfig}
\else
\usepackage[caption=false,font=footnotesize]{subfig}
\fi

\usepackage{cite}
\usepackage{bm}
\usepackage{algorithmic}
\usepackage{algorithm}
\usepackage{graphicx}
\IEEEoverridecommandlockouts

\usepackage{lettrine}

\usepackage{geometry}
\@ifundefined{showcaptionsetup}{}{%
	\PassOptionsToPackage{caption=false}{subfig}}
\usepackage{subfig}
\makeatother

\begin{document}
\setlength{\parskip}{0em}
\setlength{\abovedisplayskip}{3pt}
\setlength{\belowdisplayskip}{3pt}
\setlength{\belowdisplayshortskip}{3pt}
\newcommand{\CLASSINPUTtoptextmargin}{3pt}
\newcommand{\CLASSINPUTbottomtextmargin}{3pt}
\newcommand{\setParDis}{\setlength{\parskip}{3pt} }
\newcommand{\setParDef}{\setlength{\parskip}{0pt} }
\setlength{\floatsep}{5pt plus 2pt minus 2pt}
\setlength{\textfloatsep}{5pt plus 2pt minus 2pt}
\setlength{\intextsep}{5pt plus 2pt minus 2pt}
\title{\textcolor{black}{Distributionally Robust Chance-Constrained Optimization for Hierarchical UAV-based MEC}}
\author{
	\IEEEauthorblockN{Can Cui$^{\dagger}$, Ziye Jia$^{\dagger}$, Chao Dong$^{\dagger}$, Zhuang Ling$^{\ast}$, Jiahao You$^{\dagger}$ and Qihui Wu$^{\dagger}$\\
	\IEEEauthorblockA{$^{\dagger}$The Key Laboratory of Dynamic Cognitive System of Electromagnetic Spectrum Space, Ministry of Industry and Information Technology, Nanjing University of Aeronautics and Astronautics, Nanjing, Jiangsu, 210016, China\\
		$^{\ast}$College of Communication Engineering, Jilin University, Changchun, Jilin, 130012, China.\\
		\{cuican020619, jiaziye, dch, yjiahao, wuqihui\}@nuaa.edu.cn, lingzhuang@jlu.edu.cn}
	}
}
\maketitle
\pagestyle{empty} 

\thispagestyle{empty}
\begin{abstract}
Multi-access edge computing (MEC) is regarded as a promising technology in the sixth-generation communication. However, the antenna gain is always affected by the environment when unmanned aerial vehicles (UAVs) are served as MEC platforms, resulting in unexpected channel errors. In order to deal with the problem and reduce the power consumption in the UAV-based MEC, we jointly optimize the access scheme and power allocation in the hierarchical UAV-based MEC. Specifically, UAVs are deployed in the lower layer to collect data from ground users. Moreover, a UAV with powerful computation ability is deployed in the upper layer to assist with computing. The goal is to guarantee the quality of service and minimize the total power consumption. We consider the errors caused by various perturbations in realistic circumstances and formulate a distributionally robust chance-constrained optimization problem with an uncertainty set. The problem with chance constraints is intractable. To tackle this issue, we utilize the conditional value-at-risk method to reformulate the problem into a semidefinite programming form. Then, a joint algorithm for access scheme and power allocation is designed. Finally, we conduct simulations to demonstrate the efficiency of the proposed algorithm.
\end{abstract}

\begin{IEEEkeywords}
	Unmanned aerial vehicles, multi-access edge computing, distributionally robust optimization, conditional value-at-risk.
\end{IEEEkeywords}

\section{Introduction}
\lettrine[lines=2]{D}{ue} to the rapid development of smart devices and artificial intelligence, there exist a growing number of application data to be processed, which is an outstanding characteristic in the sixth generation of communication system \cite{VisionRequirementsTechnologyTrend6GHowTackleChallengesSystemCoverageCapacityUserDataRateMovementSpeed2020}. Besides, an emergent architecture, multi-access edge computing (MEC), becomes prevalent in recent years, which contributes to reducing the system delay and improving the quality of service (QoS) \cite{MobileEdgeComputingSurvey}. Furthermore, unmanned aerial vehicles (UAVs) are well known due to the easy deployment and flexible
movement, which attract tremendous attentions in both academics and industries. UAVs can serve as communication relay nodes, mobile base stations (BSs), as well as computing platforms for MEC \cite{intelligentservice}, \cite{LEO}. However, since there exist uncertain environmental factors such as wind, temperature, and airflow, the gains of UAVs are under uncertainty sets due to the environmental factors. For instance, the effect of wind results in the fluctuation of transmission gain within a certain range, and the extreme temperature leads to transmission gain errors. The effects of such unpredictable errors cannot be ignored in practical applications.

As illustrated that UAV-enabled MEC system can provide benefits for energy-efficient task offloading, there exist a couple of related works \cite{ComputationOffloading}. For example, \cite{MobileEdgeComputingUAVMountedCloudletOptimizationBitAllocationPathPlanning2018} studies a UAV-based mobile cloud computing system to minimize total energy consumption. In \cite{ThroughputMaximizationUAVEnabledMobileRelayingSystems2016}, Zeng $\emph{et al.}$ propose an algorithm to maximize the throughput by jointly optimizing the trajectory of the UAV swarm along with the transmit power. \cite{JointTrajectoryResourceOptimizationMECAssistedUAVsSubTHzNetworksResourcesbasedMultiAgentProximalPolicyOptimizationDRLAttentionMechanism2022a} presents a UAV-assisted MEC for task offloading which is solved with a deep reinforcement learning method. In \cite{ResponseDelayOptimizationMobileEdgeComputingEnabledUAVSwarm2020}, a response delay optimization algorithm is suggested for the deployment of UAVs as well as MEC servers. However, the uncertain errors of transmission gains are ignored or described as a certain distribution in the aforementioned works, which are not realistic in the practical environment. As above, in this paper, we consider the errors without distribution information to minimize the total power consumption in the hierarchical UAV-based MEC. Since there exist unpredictable parameters such as gain errors, which are in an uncertainty set, the proposed optimization problem is still intractable. Hence, a distibutionally robust optimization (DRO) problem considering the uncertainty of gains is further proposed to deal with this issue.

Both the stochastic programming and robust optimization can be applied to uncertain optimization problems, and the distribution information of random parameters is necessary, which cannot be obtained in some scenarios\cite{IntroductionStochasticProgramming2011}, \cite{RobustOptimization}. However, DRO can be implemented without such detailed statistical information and provide more conservative solutions \cite{OptimizationConditionalValueatRisk2000}. Besides, without distribution information on the uncertainty, the DRO problems are usually computationally-prohibitive. One approach to solving the DRO problem utilizes the conditional value-at-risk (CVaR) with limited statistical information \cite{drowban}, \cite{AoI}. Utilizing CVaR with historical data, we formulate the chance-constrained problem into a semidefinite programming (SDP) form, which can be handled with feasible solutions \cite{Distributionallyrobustjointchanceconstraintssecondordermomentinformation2013}. In this paper, we build the uncertainty set for the error parameters with historical data, enabling more practical applications. Accordingly, the system model can be formulated as a DRO problem and solved via the CVaR mechanism.
	
\textcolor{black}{In detail, we propose a hierarchical UAV-assisted MEC system to complete data processing and transmit data back to the ground station. The system is applicable to many real scenarios, such as real-time monitoring and disaster rescue for remote areas short of BS coverage \cite{Hierarchical},\cite{HAP-LEO}. Note that the UAVs are always constrained by energy, the total power consumption should be optimized, and the QoS of users also should be guaranteed. Taking into account the errors from antenna gains, we propose a chance-constrained problem with the uncertainty set and handle it by the CVaR mechanism. The main contributions of our work are summarized as follows.}
\begin{itemize}
	
	\item A hierarchical UAV-based MEC system is proposed, including the lower-layer UAVs to collect data and the upper-layer UAV for relay. Besides, both layers of UAVs are equipped with computation resources for MEC. Furthermore, the gain uncertainty from transceivers is considered for practical scenarios, and we present a corresponding system model to minimize the total power consumption under such uncertainty.
	
	\item The problem is formulated with an uncertainty set for transceiver gain errors, and to handle this issue, we propose a DRO-based mechanism, which is non-convex. Then, we approximate the chance constraints by the CVaR method and reformulate the issue into a tractable SDP form.
	
	\item To deal with the reformulated SDP problem, we present an algorithm to jointly optimize both the access scheme and power allocation. Finally, we conduct simulations to evaluate the performance of the proposed algorithm, and the results verify the effectiveness.
\end{itemize}

The rest of this paper is organized as follows. Section \ref{sec:System Model and Problem Formulation} presents the system model and corresponding problem formulation. Section \ref{sec:Approximation Reformulation and Algorithm} employs the CVaR-based mechanism to reformulate the original problem with chance constraints into a SDP form, and the corresponding algorithm is designed. Simulation results and the analyses are provided in Section \ref{sec:Simulation Results}. Finally, conclusions are drawn in Section \ref{sec:Conclusions}.
\section{System Model and Problem Fromulation\label{sec:System Model and Problem Formulation}}
\begin{figure}
	\includegraphics[width=0.99\linewidth]{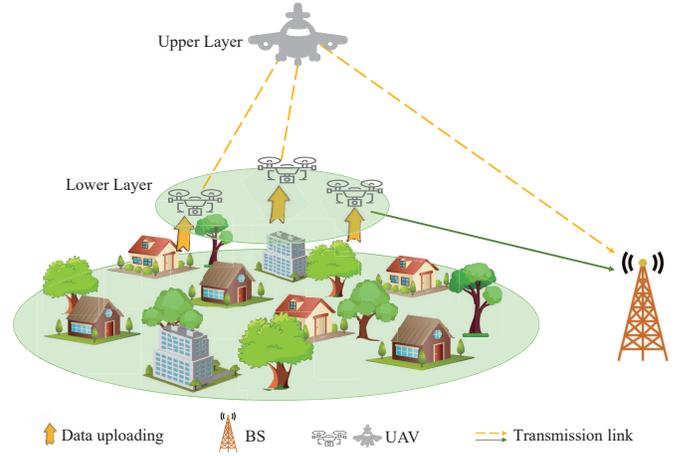}
	\caption{A hierarchical UAV-based MEC scenario.}
	\label{Example}
\end{figure}

In this paper, we consider a hierarchical UAV-assisted MEC system, in which lower-layer UAVs are deployed to collect and compute the data produced by ground users and transmit them back to the BS. Due to the limited computation resources of UAVs, a UAV with strong computing ability is deployed in the upper layer to handle the compute-intensive tasks. Lower-layer UAVs can complete the computation tasks locally and transmit the results to the BS, or transmit the data to the upper-layer UAV for further computation to reduce power consumption. Then, the upper-layer UAV transmits the processed results to the ground BS. The hierarchical UAV-based MEC scenario is shown in Fig. \ref{Example}.
\subsection{System Model}\label{A}
\subsubsection{Communication Model}
In this subsection, the communication model of UAV-to-UAV and UAV-to-BS is investigated. We utilize $\mathcal{N}=\{1,2,\ldots,i,\ldots,N\}$ to indicate the set of UAVs of the lower layer. Let \emph{$p^d_{i}$} denote the transmit power of $\emph{i-th}$ lower-layer UAV. \emph{g$_{i}$} denotes the $\emph{i-th}$ UAV transceiver antenna gain and $B_i$ indicates the bandwidth. Consequently, the data transmission rate from the $\emph{i-th}$ UAV in the lower layer to the upper-layer UAV or the BS is expressed as:
\begin{equation}
	r_i=B_i\log_{2}(1+\frac{{\gamma_ip^d_i{\left|g_i\right|}^2}}{\sigma_s^2}),\forall i\in\{1,2,\ldots,N\},
\end{equation}
where $\sigma_s^2$ is the variance of white Gaussian noise with a mean value of $0$. $\gamma_i$ is the path loss related to the distance between $\emph{i-th}$ lower-layer UAV and upper-layer UAV, or between $\emph{i-th}$ lower-layer UAV and BS.

We assume the data transmission adopts the orthogonal multiple access technology in this system, so the interference between different data can be ignored. Moreover, only one powerful UAV is considered in the upper layer.

Let $B_h$ denote the bandwidth of the upper-layer UAV. \emph{$p^d_{h}$} is the transmit power, \emph{g$_{h}$} denotes the transceiver antenna gain of the UAV hovering in the upper layer, and $\gamma_h$ represents the path loss. According to the Shannon formula, the data transmission rate of the upper layer UAV towards the BS is calculated as:
\begin{equation}
	r_h=B_h\log_{2}(1+\frac{{\gamma_hp^d_h{\left|g_h\right|}^2}}{\sigma_s^2}).
\end{equation}
\vspace{-0pt}
\vspace{-2pt}
We assume \emph{$L_i$} represents the data length transmitted by the $\emph{i-th}$ UAV. \emph{$x_i\in\{0,1\}$} is a binary indicator for computing mode. \emph{$x_i=1$} represents the $\emph{i-th}$ UAV transmits the users' data to the upper-layer UAV for further computation, and \emph{$x_i=0$} denotes the $\emph{i-th}$ UAV computes collected data locally and transmit them back to the BS.

Based on the above formulas, we can obtain the transmission delay as follows:
\begin{equation}
	t_i^d=\frac{L_i}{r_i},\forall i\in\{1,2,\ldots,N\},
\end{equation}
and 
\begin{equation}
	t_h^d=\frac{\sum_{i}^N x_i L_i}{r_h},
\end{equation}
where $t_i^d$ denotes the transmission delay of the $\emph{i-th}$ UAV in the lower layer, and $t_h^d$ is the transmission delay of the upper-layer UAV, respectively. Further, the total transmission power $p_r^d$ of UAVs in the lower layer is
\begin{equation}
	p_r^d=\sum_{i=1}^{N}p_i^d.
\end{equation}
\subsubsection{Uncertainty Model}
Take into account the transceiver antenna gain errors influenced by the environmental factors denoted as $\Delta g_i$ and $\Delta g_h$, i.e.,\begin{equation}
	g_i=\overline{g}_i+\Delta g_i,\forall i\in\{1,2,\ldots,N\},
\end{equation}
and\begin{equation}
	g_h=\overline{g}_h+\Delta g_h,
\end{equation}
where $\overline{g}_i$ and $\overline{g}_h$ indicate the theoretical antenna transmission gain of the $\emph{i-th}$ lower-layer UAV and upper-layer UAV, respectively. Similarly, $\Delta{g_i}$ and $\Delta{g_h}$ represent the antenna gain errors produced by actual environmental impacts. The estimation error $\Delta{g_i}$ and $\Delta{g_h}$ are assumed to follow an unknown distribution $\mathbb{P}$ with mean $\mathbb{E_P}(\Delta g_i)=\mu_{i}$ and $\mathbb{E_P}(\Delta g_h)=\mu_{h}$, as well as the variance $\mathbb{D_P}(\Delta g_i)=\sigma_{i}^2$ and $\mathbb{D_P}(\Delta g_h)=\sigma_{h}^2$, respectively. Uncertainty set $\mathcal{P}$ reprensents for all possible probability distributions of $\Delta{g_i}$ and $\Delta{g_h}$, i.e., $\mathbb{P}\in\mathcal{P}$.
\subsubsection{Computation Model}

Let variable \emph{$c_i$} represent the complexity of the computing tasks processed by the $\emph{i-th}$ UAV. \emph{$\eta_i$} and \emph{$\eta_h$} denote the computational power consumption coefficient of the $\emph{i-th}$ lower-layer UAV and the upper-layer UAV, respectively, which represent the power required to calculate the data of each cycle. Then we can formulate the computation power of $\emph{i-th}$ lower-layer UAV and the upper-layer UAV as:\begin{equation}
	p_i^c=\eta_ic_i,\forall i\in\{1,2,\ldots,N\},
\end{equation}
and \begin{equation}
	p_h^c=\eta_h\sum_{i=1}^N x_ic_i.
\end{equation}

With the above formulations we can obtain the total computation power consumption of UAVs in the lower layers as:
\begin{equation}
	p_r^c=\sum_{i=1}^N (1-x_i)p_i^c.
\end{equation}

\subsubsection{Power Consumption}
According to the above discussion, ignoring the power required for UAV hovering and flying, which is a constant value, the total power is expressed as:
\begin{equation}
	P_{total}=p_r^d+p_h^d+p_r^c+p_h^c.
\end{equation}

\subsection{Problem Formulation}\label{Problem Formulation}
In this part, we focus on reducing the total power in the system for a better service. The target is to minimize $P_{total}$ by jointly optimizing the UAV access scheme and power allocation when the QoS meets users' demands. Hence, the problem is formulated as
\begin{align}
\textbf{P0}:	&\min_{x_i,p_i^d,p_h^d} P_{total}, \label{objective-function}\\
	&\textrm{s.t.}\quad  \mathit{C}1: \mathbf{Pr}\{t_i^d\leq t_{i,max}^d\}\geq \alpha_1,\forall i,\label{chance-constraint-1}\\
    &\qquad \mathit{C}2: \mathbf{Pr}\{t_h^d\leq t_{h,max}^d\}\geq \alpha_2, \label{chance-constraint-2}\\
	& \qquad \mathit{C}3: 0\leq\sum_{i=1}^N x_i\leq m,\label{C3}\\
	& \qquad \mathit{C}4: x_i\in \{0,1\},\forall i,\label{C4}\\
	& \qquad \mathit{C}5: p_i^d\geq0,\forall i,\label{C5}\\
	&\qquad  \mathit{C}6: p_h^d\geq0, \label{C6}
\end{align}
where $\mathit{C}1$ and $\mathit{C}2$ are the chance constraints, indicating that the transmission delays of UAVs are constrained in a probabilistic manner. In other words, the delays should not be larger than the tolerable delay $t_{i,max}^d$ and $t_{h,max}^d$ with probabilities of $\alpha_1$ and $\alpha_2$ at a minimum, respectively. Futhermore, $\alpha_1$, $\alpha_2$ $\in$ $(0,1)$. $\mathit{C}3$ denotes that the upper-layer UAV with powerful computation ability can at most access $\mathit{m}$ UAVs at the same time. $\mathit{C}4$ denotes $\mathit{x_i}$ is a binary variable. $\mathit{C}5$ and $\mathit{C}6$ guarantee that $p_i^d$ and $p_h^d$ are non-negative continuous variables.

Due to the chance contraints (\ref{chance-constraint-1}) and (\ref{chance-constraint-2}), the transmission delay $t_i^d$ and $t_h^d$ in $\textbf{P0}$ are difficult to obtain. However, there exists one effective way to handle the problem with distributionally robust method. Specifically, let $\underset{\mathbb{P}\in\mathcal{P}}{inf}$ represent the lower bound of the probability under the probability distribution $\mathbb{P}$, and (\ref{chance-constraint-1}) and (\ref{chance-constraint-2}) can be rewritten by:
\begin{equation}
	\underset{\mathbb{P}\in\mathcal{P}}{inf}\quad\mathbf{Pr}_\mathbb{P}\{t_i^d\leq t_{i,max}^d\}\geq \alpha_1, \forall i\in\{1,2,\ldots,N\} \label{distributionally-1},
\end{equation}
and
\begin{equation}
	\underset{\mathbb{P}\in\mathcal{P}}{inf}\quad \mathbf{Pr}_\mathbb{P}\{t_h^d\leq t_{h,max}^d\}\geq \alpha_2 \label{distributionally-2},
\end{equation}
respectively, which are distributionally robust chance constraints (DRCCs).

\section{Reformulation and Algorithm\label{sec:Approximation Reformulation and Algorithm}}
Since the DRCC problems (\ref{distributionally-1}) and (\ref{distributionally-2}) are intractable with the uncertainty set, in this section, we employ the CVaR mechanism to handle the issue and then design an algorithm to obtain the final solution.
\subsection{CVaR Based Mechanism}
Generally, the value-at-risk (VaR) of a variable $u$ with the safety factor $\alpha$ is defined as the minimal value of $v$, and $u$ is no more than $v$ at a possibility of $\alpha$ \cite{valueatrisk}, i.e.,
\begin{equation}
	VaR_{\alpha}(u)=\min\{v|P(u \leq v) \geq \alpha\}.
\end{equation}

It is noted that VaR is non-convex and discontinuous. Based on VaR, we can consequently propose the definition of CVaR, which is defined as the conditional expectation of $u$ when $u \geq VaR_{\alpha}(u)$, i.e.,
\begin{equation}
	CVaR_{\alpha}(u)=E[u|u\geq VaR_{\alpha}(u)].
\end{equation}

The relationship between VaR and CVaR is shown in Fig. \ref{fig:cvar}. It is obvious that $CVaR_{\alpha}(u)\geq VaR_{\alpha}(u)$ \cite{OptimizationConditionalValueatRisk2000}. Furthermore, CVaR is a conservative approximate estimation of loss, which is robust. As is proposed in \cite{OptimizationConditionalValueatRisk2000}, for a given measurable loss function $\varphi$($\xi$): $\mathbb{R}^k \rightarrow \mathbb{R}$, CVaR under the safety factor $\alpha$ concerning the probability distribution $\mathbb {P}$ on $\mathbb{R}^k $ is expressed as:
\begin{figure}[t]
	\centering
	\includegraphics[width=1\linewidth]{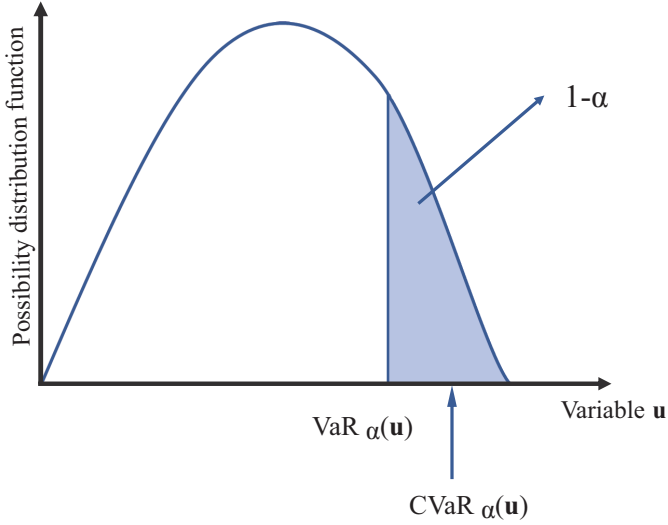}
	\caption{Relationship between VaR and CVaR.}
	\label{fig:cvar}
	\vspace{-2mm}
\end{figure}
\begin{equation}
	\mathbb{P}-CVaR_{\alpha}(\varphi(\xi))=\underset{\beta\in\mathbb{R}}{inf} \{\beta+\frac{1}{1-\alpha}\mathbb{E_P} \left[\max\left(0,\varphi(\xi)-\beta\right)\right]\}.
\end{equation}

As is shown in Fig. \ref{fig:cvar}, according to the definition, we can then obtain the following formula:
\begin{equation}
	\mathbb{P}\{\varphi(\xi)\leq\mathbb{P}-CVaR_{\alpha}(\varphi(\xi))\}\geq \alpha.
\end{equation}

As above, the CVaR constraint can be formed as:
\begin{equation}
	\begin{split}
		\underset{\mathbb{P}\in\mathcal{P}}{sup}\quad\mathbb{P}-CVaR_{\alpha}(\varphi(\xi))\leq0, \forall\mathbb{P}\in\mathcal{P} \Leftrightarrow \\
		\underset{\mathbb{P}\in\mathcal{P}}{inf}\quad\mathbb{P}\{\varphi(\xi)\leq0\}\geq\alpha.
	\end{split}\label{DRO-CVaR}
\end{equation}
Formula (\ref{DRO-CVaR}) represents that the conservative approximation of DRCC on the right side can be constituted by the CVaR constraint on the left side. In other words, CVaR represents a convex function of the random parameter. Then, we can reformulate the CVaR constraint into a tractable SDP form.

$\textbf{Lemma 1:}$ Let $\varphi(\xi)=\xi^{\top}\mathbf{\Theta}\xi+\boldsymbol{\theta}^{\top}\xi+\boldsymbol{\theta}^0$ for $\boldsymbol{\Theta}\in\mathbb{S}^k$, $\boldsymbol{\theta}\in\mathbb{R}^k$, and $\boldsymbol{\theta}^0\in\mathbf{R}$, and then, the worst-case CVaR $\underset{\mathbb{P}\in\mathcal{P}}{sup}\quad\mathbb{P}-CVaR_{\alpha}(\varphi(\xi))$ can be rewritten into a tractable SDP as follows:
\begin{equation}
	\begin{split}
		&	inf\quad\beta+\frac{1}{1-\alpha}\langle{\boldsymbol{\Omega},\boldsymbol{H}}\rangle,\\
     	\textrm{s.t.}\quad & \boldsymbol{H}\in\mathbb{S}^{k+1}, \quad\beta\in \mathbb{R},\\
		&	\boldsymbol{H}\succeq\mathbf{0},\\
		& \boldsymbol{H}-\left[\begin{matrix}
			\boldsymbol{\Theta} & \frac{1}{2}\boldsymbol{\theta}\\
			\frac{1}{2}\boldsymbol{\theta}^{\top} &\boldsymbol{\theta}^0-\beta\\
		\end{matrix}\right]\succeq\mathbf{0},
	\end{split}
\end{equation}
where $\mathbf{\Omega=\left[\begin{matrix}\Sigma+\mu\mu^{\top} & \mu\\
		\mu^{\top} & 1\\ 
	\end{matrix}\right]}$, $\Sigma\in\mathbb{S}^k$ is the covariance of the random vector $\xi$, and $\mu\in\mathbb{R}^k$ is the mean matrix. Furthermore, $\boldsymbol{H}$ is an auxiliary matrix and $\beta$ is an auxiliary variable. $\langle{\boldsymbol{\Omega},\boldsymbol{H}}\rangle=\mathrm{Tr} (\boldsymbol{\Omega}\boldsymbol{H})$ represents the trace scalar product. $\boldsymbol{H}\succeq\mathbf{0}$ represents that the matrix $\boldsymbol{H}$ is positive semidefinite\cite{Distributionallyrobustjointchanceconstraintssecondordermomentinformation2013}.

Therefore, according to Lemma 1, we transform the DRO into the form of SDP by the CVaR mechanism, which is a tractable and computationally efficient approximation technique. This form can be regarded as an extension of linear programming, and the effectiveness of this solution is proved in previous works \cite{sdp}.
\subsection{Problem Reformulation}
\setParDis
As for (\ref{distributionally-1}), a DRCC problem, the first-order Taylor expansion of the loss function is
\begin{equation}
	\varphi(\xi_1)=L_i\sigma_s^2-B_i\gamma_ip^d_it_{i,max}^d\xi_1^2, \label{loss-function-1}
\end{equation}
where $\xi_1$ denotes the random parameter $\left|g_i\right|$.

CVaR can construct a convex approximation for the chance constraints. According to Lemma 1, we transform (\ref{loss-function-1}) into a SDP problem, i.e.,
\begin{equation}
		\begin{split}
		&	inf\quad\beta_1+\frac{1}{1-\alpha_1}\langle{\boldsymbol{\Omega_1},\boldsymbol{H_1}}\rangle\leq0,\\
		\textrm{s.t.}\quad& \boldsymbol{H_1}\in\mathbb{S}^{2}, \quad\beta_1\in \mathbb{R},\\
		&	\boldsymbol{H_1}\succeq\mathbf{0},\\
		& \boldsymbol{H_1}-\left[\begin{matrix}
			-B_i\gamma_ip_i^dt_{i,max}^d & 0\\
			0 &L_i\sigma_s^2-\beta_1\\
		\end{matrix}\right]\succeq\mathbf{0},
	\end{split} \label{SDP-1}
\end{equation}
where $\boldsymbol{H_1}$ and $\beta_1$ are both auxiliary variables. Let $\mu_{g_i}=\overline{g}_i+\mu_i$ present the mean of $g_i$, we have
\begin{equation}
	\Omega_1=\left[\begin{matrix}
		 \sigma_{i}^2+\mu_{g_i}\mu_{g_i}^{\top} & \mu_{g_i} \\
		 \mu_{g_i}^{\top} & 1\\
	\end{matrix}\right].
\end{equation}

Likewise, let $\xi_2$ denote the random parameter $\left|g_h\right|$, and the loss function in (\ref{distributionally-2}) can be rewritten as:
\begin{equation}
	\varphi(\xi_2)=\sum_{i=1}^N x_iL_i\sigma_s^2-B_h\gamma_hp^d_ht_{h,max}^d\xi_2^2,
\end{equation}
which can be further presented in a SDP form, i.e.,
\begin{equation}
	\begin{split}
		&	inf\quad\beta_2+\frac{1}{1-\alpha_2}\langle{\boldsymbol{\Omega_2},\boldsymbol{H_2}}\rangle\leq0,\\
		\textrm{s.t.}\quad& \boldsymbol{H_2}\in\mathbb{S}^{2}, \quad\beta_2\in \mathbb{R},\\
		&	\boldsymbol{H_2}\succeq\mathbf{0},\\
		& \boldsymbol{H_2}-\left[\begin{matrix}
			-B_h\gamma_hp^d_ht_{h,max}^d & 0\\
			0 &\sum_{i=1}^Nx_iL_i\sigma_s^2-\beta_2\\
		\end{matrix}\right]\succeq\mathbf{0}.
	\end{split} \label{SDP-2}
\end{equation}
Similarly, $\boldsymbol{H_2}$ is the auxiliary matrix, and $\beta_2$ is the auxiliary variable for the CVaR constraint. $\mu_{g_h}=\overline{g}_h+\mu_h$ is the mean value of $g_h$. Then we obtain
\begin{equation}
	\Omega_2=\left[\begin{matrix}
		\sigma_{h}^2+\mu_{g_h}\mu_{g_h}^{\top} & \mu_{g_h} \\
		\mu_{g_h}^{\top} & 1\\
	\end{matrix}\right].
\end{equation}

Finally, based on above discussions, the chance-constrained problem (\ref{objective-function}) under the uncertainty set can be reformulated into a SDP form, i.e.,
\begin{equation}
	\begin{split}
		\textbf{P1}:\quad&\min_{\substack{x_i,p^d_i,p^d_h,\\\boldsymbol{H_1},\boldsymbol{H_2},\beta_1,\beta_2}}P_{total},\\
		&\mathbf{\textrm{s.t.}}\quad(\ref{C3})-(\ref{C6}), (\ref{SDP-1}), (\ref{SDP-2}),\\
		&\forall i\in\{1,2,\ldots,N\},
	\end{split}\nonumber\quad
\end{equation}
which is still non-convex with the binary variable $x_i$, and continuous variables $p^d_i$ and $p^d_h$. The problem is in the form of mixed integer non-linear programming. To tackle with this problem, we design a joint optimization algorithm on access scheme and power allocation, as shown in Algorithm \ref{alg:t}.
\setParDef
\subsection{Algorithm Design}
It is observed that due to the two decision variables $x$ and $p^d$, problem $\textbf{P1}$ is still non-convex. To solve this problem, we divide it into two subproblems. The first subproblem is concerning the access scheme $x$, and the second subproblem is related to the transmission power allocation plan $p^d$.

As shown in step 2 of Algorithm \ref{alg:t}, in the first subproblem, an initial value $p^d$ is assumed to minimize the computation power consumption and ensure the access scheme, i.e., $x$. In other words, the lower-layer UAVs make a choice to complete the computation work locally or transform it to the upper-layer UAV based on the task complexity. The problem with integers is solved by MOSEK. Then, in the second subproblem, according to the access scheme $x^*$ that is obtained in step 2, we can further substitute it to solve the transmission power allocation program. Specifically, the subproblem can be tackled by fixing the access scheme, i.e., $x^*$. Then, in step 3 of Algorithm \ref{alg:t}, the transmission power is allocated and variables $p^d_i$ as well as $p^d_h$ are obtained. Repeat the process until the solutions of two subproblems converge. The final result, i.e., the total power consumption $P_{total}$ in the hierarchical UAV-assisted MEC system, can be figured out as $\boldsymbol{optval}$.
\begin{algorithm}[t]
	\caption{Joint Optimization Algorithm on Access Sheme and Power Allocation.\label{alg:t}}
	\begin{algorithmic}[1]
		\REQUIRE Network parameters, $\boldsymbol{c_i}$, $\boldsymbol{L_i}$.
		\ENSURE $\boldsymbol{x_i}$, $\boldsymbol{p^d_i}$, $\boldsymbol{p^d_h}$ and minimum system power consumption $\boldsymbol{optval}$.
		\REPEAT\nonumber
		\STATE Solve the subproblem of access-scheme on $\boldsymbol{x_i}$ to obtain $\boldsymbol{x_i^*}$ and the power consumption $\boldsymbol{\tau_1}$. $\boldsymbol{x_i}$ $\rightarrow$ $\boldsymbol{x_i^*}$.
		\STATE Solve the subproblem of power $\boldsymbol{p^d}$ to obtain $\boldsymbol{p^{d*}}$ and the power consumption $\boldsymbol{\tau_2}$. $\boldsymbol{p^d}$ $\rightarrow$ $\boldsymbol{p^{d*}}$.
		\UNTIL{Results converge.}\nonumber
	    \STATE Obtain the final total power $\boldsymbol{optval}$.
	\end{algorithmic}
\end{algorithm}
\section{Simulation Results\label{sec:Simulation Results}}
Numeral simulations are conducted to evaluate the performance of the proposed algorithm. Briefly speaking, the data lengths are assumed to range from $40kbits$ to $60kbits$. $B_i$ and $B_h$ are assumed as $10MHz$ and $10MHz$, respectively. The task complexities range from $40kcycles$ to $100kcycles$. The CPU computational power consumption coefficient $\eta_i$ is $5\times10^{-6}cycles/W$, and $\eta_h$ is $10^{-6}cycles/W$. Antenna gains $\left|g_i\right|$ and $\left|g_h\right|$ are $5$. The mean value of errors is $0$ and the variance is $0.01$ for all. Furthermore, noise $\sigma_{s}^2$ is assumed as $10^{-12}W$. The safety factors are $\alpha_1=0.95$ and $\alpha_2=0.95$.

\begin{figure}[t]
	\centering
	\includegraphics[width=1.0\linewidth]{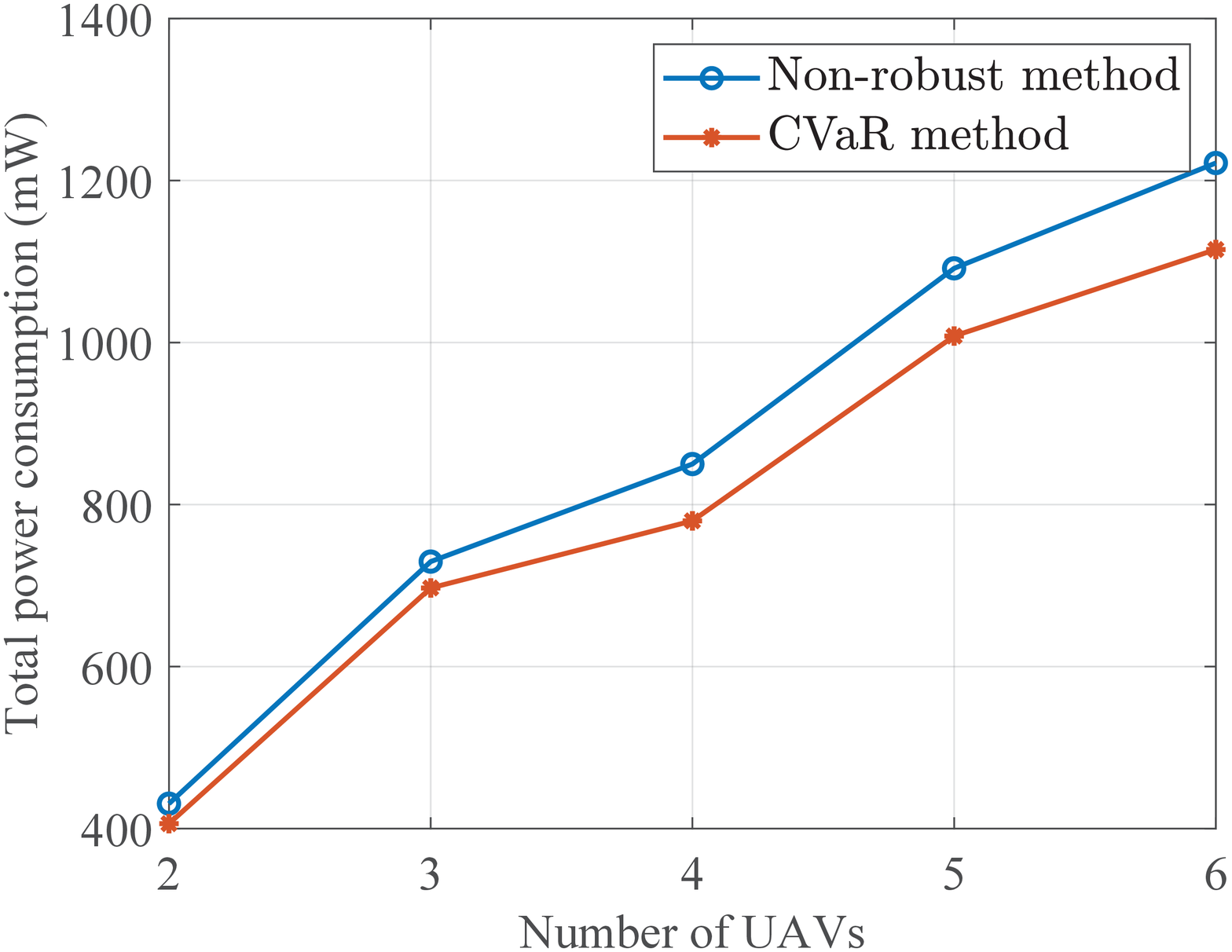}
	\caption{System power $\emph{v.s.}$ the number of lower-layer UAVs for different methods.}
	\label{duibi}
\end{figure}

To verify the effectiveness and feasibility, we compare the results of the CVaR method and non-robust method without uncertainty errors $\Delta{g_i}$ and $\Delta{g_h}$. In detail, Fig. \ref{duibi} shows the performance of two methods in total power $\emph{v.s.}$ the number of lower-layer UAVs. As the number of lower-layer UAVs increases, the total power consumption of the MEC system increase as well. It can be explained that as the scale of the whole problem becomes larger, the users' data collected are larger to be computed and transmitted. Comparing the results of the two methods, it is observed the power consumption in the CVaR method is smaller than that in the non-robust method. It is explained that there is a tolerance in the CVaR method, so that the transmission power of each UAV can be reduced within the tolerant range. Moreover, different from the non-robust method, which optimizes the problems under the perfect assumption which is apparently inconsistent with the real circumstance, the designed algorithm takes into account the uncertainties in reality and is applicable to the real situation.

\begin{figure}
	\centering
	\includegraphics[width=1.0\linewidth]{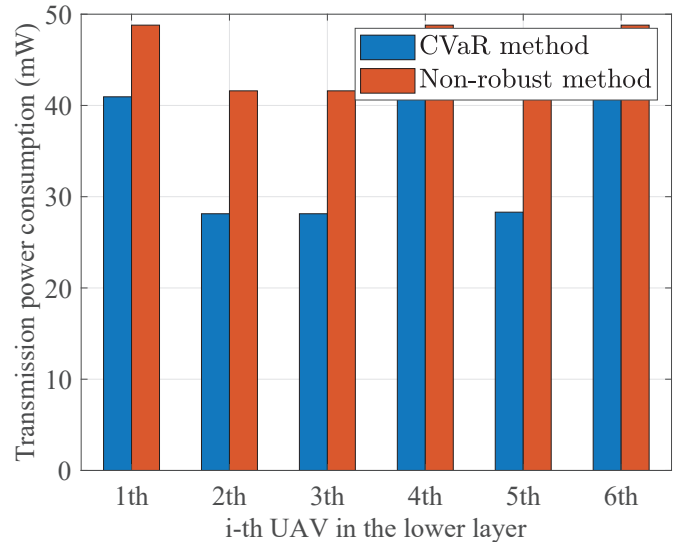}
	\caption{Individial transmission power of the i-th UAV in the lower layer for different methods.}
	\label{ith-UAV}
\end{figure}
Fig. \ref{ith-UAV} shows the individual transmission power of each lower-layer UAV when there exist six UAVs in the lower layer. We compare the result with the non-robust method. For tasks varying in data length, it demonstrates that the CVaR method performs better in transmission power consumption. The conclusion can be drawn that the proposed method, which is considered in a more realistic world, can obtain a better result than that in the ideal situation. Specifically, considering the errors between the real situation and the ideal situation, the designed algorithm only needs to obtain the mean and variance of the gain errors from the historical data, instead of obtaining the true distribution or probability function of the errors. The experimental results verify the effectiveness of the proposed joint optimization algorithm on access scheme and power allocation.

\begin{figure}[t]
	\centering
	\includegraphics[width=1.0\linewidth]{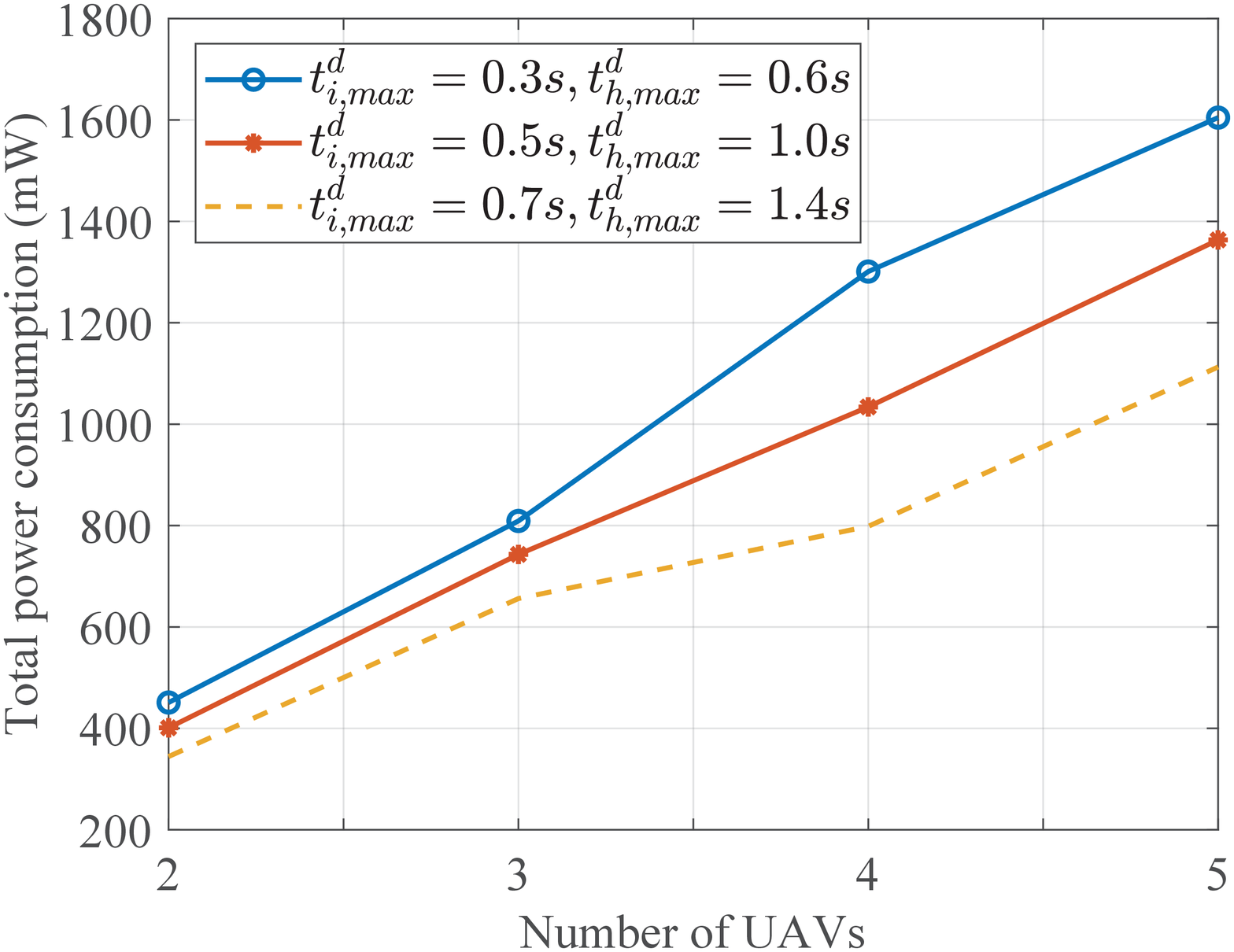}
	\caption{System power $\emph{v.s.}$ the number of lower-layer UAVs for different maximum transmission delays.}
	\label{n-p}
\end{figure}

In Fig. \ref{n-p}, we investigate the performance of total power consumption, i,e., the corresponding system power $\emph{v.s.}$ lower-layer UAV numbers for different maximum transmission delays. It depicts the influence of the maximum tolerance on the total power. With the increment of maximum delay to be satisfied, the transmission power consumption reduces. It is accounted for the fact that transmission rates become lower for all UAVs when the tolerance transmission delay increases.

\section{Conclusions\label{sec:Conclusions}}
In this paper, we study a hierarchical UAV-assisted MEC scenario to optimize the power consumption. To tackle with the potential uncertainty impacted by environmental factors, which is not ignorable in practical circumstances, a DRO problem based on the uncertainty set is proposed. Then, with the CVaR mechanism, the original problem is reformulated into a SDP form. To jointly optimize the reformulation on the access scheme and power allocation, we further design an algorithm to handle it and obtain the final solution. By conducting simulation experiments, the robustness and feasibility of the proposed algorithm compared with the non-robust method are verified.
\bibliographystyle{IEEEtran}
\bibliography{reference.bib}
\end{document}